# Resonant inter-mode second harmonic generation by backward spin waves in YIG nano-waveguides

K.O. Nikolaev[1,‡], S. R. Lake[2,‡], B. Das Mohapatra[2,‡], G. Schmidt[2,3], S. O. Demokritov[1], and V. E. Demidov[1*]

[1]*Institute of Applied Physics, University of Muenster, 48149 Muenster, Germany*

[2]*Institut für Physik, Martin-Luther-Universität Halle-Wittenberg, 06120 Halle, Germany*

[3]*Interdisziplinäres Zentrum für Materialwissenschaften, Martin-Luther-Universität Halle-Wittenberg, 06120 Halle, Germany*

We experimentally study nonlinear generation of the second harmonic by backward volume spin waves propagating in microscopic magnonic waveguides fabricated from a low-loss magnetic insulator with a thickness of several tens of nanometers. We show that highly efficient resonant second harmonic generation is possible in the inter-mode regime at microwave powers of the order of $10^{-4}$ W. In contrast to previously observed second harmonic generation processes, the generation by backward waves is characterized by the nonlinearly generated waves propagating opposite to the initial waves, and can be realized at zero bias magnetic field.

‡ : authors that contributed equally to the work

*Corresponding author, e-mail: demidov@uni-muenster.de

One of the most important and practical advantages of large-amplitude spin-wave dynamics, which is not available in the small-amplitude linear propagation regime, is the possibility to nonlinearly generate spin waves with frequencies and wavelengths different from those of the initially excited wave[1-16]. This possibility can be used, for example, to excite waves that are difficult to generate directly using a linear excitation mechanism due to their short wavelengths[9,14,16], to perform frequency up- and down-conversion of signals[2-8,13], as well as for amplitude- and phase-shaping of spin-wave pulses[1,10]. In addition to these traditional signal-processing applications, nonlinear coupling of spin-wave modes at different frequencies has also been shown to enable the implementation of systems capable of neuromorphic-like computing[12].

Processes leading to nonlinear generation of additional spectral components can be described in terms of three- and four-magnon interactions[17]. For these interactions to be efficient, the conservation of energy and momentum for interacting magnons must be satisfied simultaneously. Due to this strict constraint, the peculiar characteristics of magnon dispersion are essential to achieve efficient resonant nonlinear generation. One of the important examples is the generation of the second harmonic by propagating spin waves[2-8]. Because of the nonlinear dependence of the frequency of spin waves on their wavenumber, it is generally difficult to achieve resonant conditions for magnons with frequencies that differ by a factor of two. Recently it was shown that this can be accomplished using an inter-mode interaction process in microscopic waveguides with an engineered dispersion spectrum[14].

In this Letter, we experimentally study the inter-mode resonant second harmonic generation for spin waves propagating in yttrium-iron-garnet (YIG) nano-waveguides, where the static magnetization is oriented parallel to the waveguide axis. This configuration is particularly promising for technical applications because it establishes a way for magnonic circuits to operate under small[18] or even zero[19,20] bias magnetic field. We show that, in YIG waveguides with a thickness of several tens of nanometers, efficient resonant second-harmonic generation can be achieved for the above geometry, corresponding to backward-volume spin waves. Due to the backward dispersion of these waves, the second-harmonic wave propagates in the direction opposite to the direction of the initial wave. This results in a maximization of the intensity of the second-harmonic wave near the spatial position where the initial wave is excited, in stark contrast with waves with forward dispersion that exhibit a gradual increase in the intensity of the second harmonic along the propagation distance. This feature can be used to control the propagation direction of the second-harmonic wave by simply changing the direction of the bias magnetic field. We also show that by varying the thickness of the waveguide, one can achieve resonant second-harmonic generation conditions at zero bias magnetic field, which paves the way for implementation of nonlinear magnonic devices and circuits that do not require a bias magnetic field for their operation.

In Fig. 1, we show the schematics of the experiment. We study the generation of the second harmonic by spin waves propagating in a 500 nm wide waveguide fabricated by electron-beam lithography and lift-off technique from an 80 nm thick YIG film deposited using pulsed-laser deposition[21]. The spin waves are excited using an inductive Au antenna that is 500 nm wide and 200 nm thick. The waveguide is magnetized by a static magnetic field $H$ applied parallel to its axis. Therefore, the excited spin waves propagate along the direction of the static magnetization. This corresponds to the case of the so-called backward-volume spin waves (BVSW)[17] known for their peculiar dispersion relation, which will be discussed below.

We study the propagation of the excited spin waves and the generation of the second harmonic using micro-focus Brillouin light scattering (BLS) spectroscopy[22]. We focus the probing laser light with a wavelength of 473 nm into a diffraction-limited spot on the surface of the YIG film (see Fig. 1) and analyze the modulation of this light due to its inelastic scattering from spin waves. The BLS intensity is proportional to the intensity of the spin wave at the position of the laser spot, which enables one to analyze the propagation of spin waves with sub-micrometer spatial resolution. Thanks to the frequency resolution of the technique, the initial spin wave and the nonlinearly generated second harmonic can be observed independently. Additionally, by using the interference of scattered light with light modulated at the frequency of spin waves, we can also resolve the phase of spin waves. This allows us to directly measure the dispersion relations of the waves under study[22].

Figure 2 shows the dispersion relations obtained from the experiment (diamonds) and from micromagnetic simulations (solid curves) at $H$ = 300 Oe. The calculations are performed using the simulation package mumax3[23] and the method described in detail in Ref. 24. In the calculations, we use the standard for YIG saturation magnetization, $4\pi M_s$ = 1750 G and the nominal YIG thickness of 80 nm. The width of the waveguide $w$ is used as an adjustable parameter. The best agreement with the experimental data is achieved for $w$ = 480 nm, which is very close to the nominal value of 500 nm. The dispersion relation $f_0(k)$ corresponds to the fundamental mode of the waveguide characterized by a uniform distribution of the dynamic magnetization through the thickness of the film and the relation $f_1(k)$ corresponds to the first-order thickness mode characterized by a non-uniform thickness distribution (see the insets in Fig. 2).

As seen from Fig. 2, the dispersion curve of the fundamental mode exhibits behaviors typical for BVSW – the frequency of the wave decreases with increasing wavenumber $k$ (often referred to as backward dispersion). This dispersion leads to opposite signs of the phase velocity $v_p = 2\pi \frac{f}{k}$ and the group velocity $v_g = 2\pi \frac{df}{dk}$. In practice, this means that a wave propagating away from the antenna in the positive direction of the $z$-axis has a wavevector oriented in the negative direction, i.e., towards the antenna. In most cases, this feature of BVSW does not play a decisive role. However, as will be shown below, it has a crucial impact on the relevant process of resonant second harmonic generation.

In terms of magnon interactions, the process of second-harmonic generation can be considered as the confluence of two magnons of the initially excited spin wave into a magnon of doubled energy (frequency), as required by the energy conservation law. Simultaneously, to conserve the linear momentum, the wavevector of the new magnon must be twice as large as the wavevector of the initial magnons. Accordingly, efficient generation of the second harmonic can be achieved only if the magnon states with the combination of wavenumber and frequency $(k, f)$ and $(2k, 2f)$ belong to the resonant spectrum of spin waves.

To find out whether this condition can be satisfied for BVSW, we plot in Fig. 2 function $2f_0(2k)$ (dashed curve) and find its intersections with the dispersion curves of spin-wave modes. As seen from Fig. 2, the curve $2f_0(2k)$ does not intersect the dispersion curve of the fundamental mode $f_0(k)$ anywhere. However, it intersects the dispersion curve of the first-order thickness mode $f_1(k)$ at the point marked by a solid circle. This indicates that the resonant phase-matched generation of the second harmonic by BVSW is possible in the inter-mode regime (dashed arrow in Fig. 2): a fundamental-mode spin wave with frequency $f_r$ and wavenumber $k_r$ (open circle in Fig. 2) can resonantly generate a second-harmonic spin wave with frequency $2f_r$ and wavenumber $2k_r$, which belongs to the first-order thickness mode.

We emphasize that in the case of BVSW, the resonant second harmonic generation has rather unusual features. As seen from Fig. 2, in contrast to the dispersion curve of the fundamental mode $f_0(k)$, the dispersion curve $f_1(k)$ exhibits a positive slope. Due to the positive slope, for this mode, the direction of the group velocity coincides with the direction of the phase velocity, i.e., the direction of the energy flow coincides with the direction of the wavevector. Since, as discussed above, the initial BVSW transferring energy in the positive $z$-direction possesses a negative wavenumber $k_r < 0$, the second harmonic possesses a negative wavenumber $2k_r < 0$, as well. For this wave, this corresponds to an energy transfer in the negative direction of the $z$ axis, i.e., towards the antenna, as schematically shown in Fig. 1.

Figures 3(a) and 3(b) present experimental evidence for the resonant second harmonic generation described above. In these experiments, we place the probing laser spot on the axis of the waveguide at a distance $z = 1$ μm from the center of the antenna and measure the BLS spectra as a function of the excitation frequency $f_e$. The power of the excitation signal is fixed at $P_e$ = 0.05 mW. As seen from Fig. 3(a), we detect linear excitation of spin waves ($f = f_e$) in the frequency range of 2.4 – 2.8 GHz, which corresponds to the frequency band of BVSW (Fig. 2). In addition to the signal at frequency $f = f_e$, we also observe an intense spectral peak at $f = 2f_e$, when the excitation frequency is close to 2.57 GHz (horizontal dashed line in Fig. 3(a)). This frequency corresponds to the spectral state ($k_r$, $f_r$) in Fig. 2, for which the conservation conditions are satisfied. Note that at all other frequencies, the BLS intensity at $f = 2f_e$ remains negligible. This indicates that the generation of the second harmonic has a well-pronounced resonant character and becomes highly efficient only when the conservation conditions are met. Figure 3(b) shows the BLS spectrum recorded at $f_e$ = 2.57 GHz (section of Fig. 3(a) along the horizontal dashed line). As seen from these data, the detected BLS intensity from the second harmonic significantly exceeds the BLS intensity from the initial wave, which reflects the high efficiency of the resonant process.

Figure 3(c) shows the BLS intensities from the second harmonic and the initial wave as a function of the excitation power $P_e$ recorded at $f_e$ = 2.57 GHz. These data show that the ratio of the two intensities increases with increasing power, reaching more than a factor of five at $P_e$ = 0.2 mW. In other words, an increase in $P_e$ leads to an increase in the efficiency of the second harmonic generation process. As seen from Fig. 3(c), the intensity of the second harmonic starts to saturate at $P_e$ > 0.15 mW. This is likely due to an onset of parasitic nonlinear damping associated with four-magnon scattering of the initial wave into short-wavelength spin-wave states that cannot be detected by BLS[25-27]. This assumption is consistent with the observed power dependence of the intensity of the initial wave recorded near the antenna (squares in Fig. 3(c)). The dependence is linear at small powers and starts to saturate at $P_e$ > 0.15 mW. This indicates that at $P_e$ > 0.15 mW, the efficiency of the excitation by the antenna starts to decrease, which is a clear signature of the onset of nonlinear damping. Note that, in contrast to four-magnon scattering processes, which become active when the intensity of the spin wave exceeds a certain threshold, the second harmonic generation is a threshold-less phenomenon, which is present at any intensity. Therefore, by setting the excitation power slightly below the threshold of the nonlinear scattering, one can achieve optimal energy efficiency of the process.

We now turn to the analysis of the spatial dependences of the intensities of the initial wave and the second harmonic. Figure 4 shows these dependences recorded at $f_e$ = 2.57 GHz and $P_e$ = 0.05 mW. As seen from these data, the initial wave shows a well-defined exponential decrease with the distance from the antenna ~ $\exp(-2z/\xi)$, where $\xi$ is the propagation length (note the logarithmic scale of the vertical axis). This exponential decay is observed over the entire range $P_e$ = 0.01 – 0.2 mW, although the propagation length strongly decreases with the increase in $P_e$ (Fig. 4(b)). Note that this decrease is pronounced even at $P_e$ < 0.15 mW, i.e., it is not dominated by the nonlinear damping due to four-magnon scattering. We attribute these behaviors to an increase in the efficiency of the second-harmonic generation (Fig. 3(c)), which results in faster energy transfer from the initial wave to the second harmonic and leads to faster spatial decay of the former.

Similar to the initial wave, the intensity of the second harmonic also exhibits a decrease with distance from the antenna (albeit it is clearly non-exponential). We emphasize that this is a non-trivial observation. For optical waves[28] as well as for spin waves possessing forward dispersion[14], the intensity of the second harmonic typically increases with the propagation distance, which is a natural result of the continuous resonant energy transfer between the initial wave and the co-propagating second harmonic. However, as discussed above, in the case of BVSW, the direction of the energy flow for the second harmonic is opposite to that of the initial wave. As a result of this opposite flow of energy, the intensity of the second harmonic maximizes close to the input antenna. This interpretation is also supported by the intensity oscillations observed near the antenna ($z$ = 1-5 μm). We associate these oscillations with the interference of two counterpropagating second-harmonic waves generated to the left and right of the antenna.

Finally, we analyze the controllability of the second harmonic generation by the static magnetic field. We perform measurement similar to those characterized in Fig. 3(a) at $H$ varying from 200 to 400 Oe and determine the resonant frequency $f_r$ for each field value. Additionally, from phase-resolved measurements at $f_r$, we find the wavelength of BVSW corresponding to the resonant conditions. As seen from Fig. 5(a), the resonant frequency increases with the increase in $H$. This is expected, since the entire dispersion spectrum (Fig. 2) shifts towards higher frequencies. Simultaneously, an increase in $H$ leads to a decrease in the resonant wavelength, which is caused by a relative shift of the $2f_0(2k)$ curve with respect to the $f_1(k)$ curve. Due to this shift, the intersection point (filled circle in Fig. 2) moves towards larger $k$ (smaller wavelengths), as the field increases. Note, that the resonant condition can be satisfied at any field $H > H_{min}$, where $H_{min}$ is determined by the condition $2f_0(0) = f_1(0)$ (intersection at $k = 0$). Figure 5(b) shows the field dependences of $2f_0(0)$ and $f_1(0)$ obtained from micromagnetic simulations. These data show $H_{min} = 150$ Oe for the parameters used in our experiment. Because the frequencies of the first-order thickness mode strongly depend on the waveguide thickness, $H_{min}$ can be further reduced by increasing the thickness, as shown in Fig. 5(c). Note that for thicknesses > 89 nm, resonant generation of the second harmonic can also be realized for BVSW propagating at zero bias magnetic field[20].

In conclusion, we have experimentally demonstrated highly efficient resonant generation of second harmonic by backward volume spin waves propagating in YIG nano-waveguides magnetized parallel to the axis. We have shown that the peculiar dispersion properties of these waves result in unusual features, such as the opposite propagation of the initial wave and the second harmonic wave. Together with the possibility to achieve resonant processes at zero bias magnetic field, this feature paves the way for the realization of highly flexible nonlinear spin-wave devices and circuits for advanced signal processing.

This work was supported by the Deutsche Forschungsgemeinschaft (DFG, German Research Foundation) – project number 529812702.

**REFERENCES**

1. M. Wu and B. A. Kalinikos, "Coupled Modulational Instability of Copropagating Spin Waves in Magnetic Thin Films," Phys. Rev. Lett. **101**, 027206 (2008).

2. V. E. Demidov, M. P. Kostylev, K. Rott, P. Krzysteczko, G. Reiss, and S. O. Demokritov, "Generation of the second harmonic by spin waves propagating in microscopic stripes," Phys. Rev. B **83**, 054408 (2011).

3. J. Marsh, V. Zagorodnii, Z. Celinski, and R. E. Camley, "Nonlinearly generated harmonic signals in ultra-small waveguides with magnetic films: Tunable enhancements of 2nd and 4th harmonics," Appl. Phys. Lett. **100**, 102404 (2012).

4. O. Rousseau, M. Yamada, K. Miura, S. Ogawa, and Y. Otani, "Propagation of nonlinearly generated harmonic spin waves in microscopic stripes," J. Appl. Phys. **115**, 053914 (2014).

5. S. Wang, J. Ding, X. Guan, M. B. Jungfleisch, Z. Zhang, X. Wang, W. Gu, Y. Zhu, J. E. Pearson, X. Cheng, A. Hoffmann, and X. Miao, "Linear and nonlinear spin-wave dynamics in ultralow-damping microstructured Co2FeAl Heusler waveguide," Appl. Phys. Lett. **113**, 232404 (2018).

6. H. J. J. Liu, G. A. Riley, C. L. Ordóñez-Romero, B. A. Kalinikos, and K. S. Buchanan, "Time-resolved study of nonlinear three-magnon processes in yttrium iron garnet films," Phys. Rev. B **99**, 024429 (2019).

7. D. R. Rodrigues, J. Nothhelfer, M. Mohseni, R. Knapman, P. Pirro, and K. Everschor-Sitte, "Nonlinear Dynamics of Topological Ferromagnetic Textures for Frequency Multiplication," Phys. Rev. Applied **16**, 014020 (2021).

8. F. Groß, M. Weigand, A. Gangwar, M. Werner, G. Schütz, E. J. Goering, C. H. Back, and J. Gräfe, "Imaging magnonic frequency multiplication in nanostructured antidot lattices," Phys. Rev. B **106**, 014426 (2022).

9. R. Dreyer, A. F. Schäffer, H. G. Bauer, N. Liebing, J. Berakdar, and G. Woltersdorf, "Imaging and phase-locking of non-linear spin waves," Nat. Commun. **13**, 4939 (2022).

10. H. Merbouche, B. Divinskiy, K. O. Nikolaev, C. Kaspar, W. H. P. Pernice, D. Gouéré, R. Lebrun, V. Cros, J. Ben Youssef, P. Bortolotti, A. Anane, S. O. Demokritov, and V. E. Demidov, "Giant nonlinear self-phase modulation of large-amplitude spin waves in microscopic YIG waveguides," Sci. Rep. **12**, 7246 (2022).

11. T. Hache, L. Körber, T. Hula, K. Lenz, A. Kákay, O. Hellwig, J. Lindner, J. Fassbender, and H. Schultheiss, "Control of four-magnon scattering by pure spin current in a magnonic waveguide," Phys. Rev. Applied **20**, 014062 (2023).

12. L. Körber, C. Heins, T. Hula, J.-V. Kim, S. Thlang, H. Schultheiss, J. Fassbender, and K. Schultheiss, "Pattern recognition in reciprocal space with a magnon-scattering reservoir," Nat. Commun. **14**, 3954 (2023).

13. M. G. Copus, T. Hula, C. Heins, L. Flacke, M. Weiler, K. Schultheiss, H. Schultheiss, and R. E. Camley, "Generation of localized, half-frequency spin waves in micron sized ferromagnetic stripes: Experiments and simulations," Appl. Phys. Lett. **124**, 192401 (2024).

14. K. O. Nikolaev, S. R. Lake, G. Schmidt, S. O. Demokritov, and V. E. Demidov, "Resonant generation of propagating second-harmonic spin waves in nano-waveguides," Nat. Commun. 15, 1827 (2024).

15. K. An, M. Xu, A. Mucchietto, C. Kim, K.-W. Moon, C. Hwang, and D. Grundler, "Emergent coherent modes in nonlinear magnonic waveguides detected at ultrahigh frequency resolution," Nat. Comm. **15**, 7302 (2024).

16. K. O. Nikolaev, B. Das Mohapatra, G. Schmidt, S. O. Demokritov, and V. E. Demidov, "Spatially extended nonlinear generation of short-wavelength spin waves in yttrium iron garnet nanowaveguides," Phys. Rev. Applied **22**, 044083 (2024).

17. A. G. Gurevich and G. A. Melkov, *Magnetization Oscillations and Waves* (CRC, New York, 1996).

18. Q. Wang, M. Kewenig, M. Schneider, R. Verba, F. Kohl, B. Heinz, M. Geilen, M. Mohseni, B. Lägel, F. Ciubotaru, C. Adelmann, C. Dubs, S. D. Cotofana, O. V. Dobrovolskiy, T. Brächer, P. Pirro, and A. V. Chumak, "A magnonic directional coupler for integrated magnonic half-adders," Nat. Electr. **3**, 765–774 (2020).

19. Q. Wang, P. Pirro, R. Verba, A. Slavin, B. Hillebrands, A.V. Chumak, "Reconfigurable nanoscale spin-wave directional coupler," Sci. Adv. **4**, e1701517 (2018).

20. K. O. Nikolaev, S. R. Lake, G. Schmidt, S. O. Demokritov, and V. E. Demidov, "Zero-Field Spin Waves in YIG Nanowaveguides," Nano Lett. **23**, 8719–8724 (2023).

21. C. Hauser, T. Richter, N. Homonnay, C. Eisenschmidt, M. Qaid, H. Deniz, D. Hesse, M. Sawicki, S. G. Ebbinghaus, and G. Schmidt, "Yttrium iron garnet thin films with very low damping obtained by recrystallization of amorphous material," Sci. Rep. **6**, 20827 (2016).

22. V. E. Demidov and S. O. Demokritov, "Magnonic waveguides studied by micro-focus Brillouin light scattering," IEEE Trans. Mag. **51**, 0800215 (2015).

23. A. Vansteenkiste, J. Leliaert, M. Dvornik, M. Helsen, F. Garcia-Sanchez, and B. Van Waeyenberge, "The design and verification of MuMax3," AIP Adv. **4**, 107133 (2014).

24. H. Merbouche, B. Divinski, D. Gouéré, R. Lebrun, A. El Kanj, V. Cros, P. Bortolotti, A. Anane, S. O. Demokritov, and V. E. Demidov, "True amplification of spin waves in magnonic nano-waveguides," Nat. Commun. **15**, 1560 (2024).

25. M. M. Scott, C. E. Patton, M. P. Kostylev, and B. A. Kalinikos, "Nonlinear damping of high-power magnetostatic waves in yttrium–iron–garnet films," J. Appl. Phys. **95**, 6294 (2004).

26. T. Hula, K. Schultheiss, A. Buzdakov, L. Körber, M. Bejarano, L. Flacke, L. Liensberger, M. Weiler, J. M. Shaw, H. T. Nembach, J. Fassbender, and H. Schultheiss, "Nonlinear losses in magnon transport due to four-magnon scattering," Appl. Phys. Lett. **117**, 042404 (2020).

27. S. R. Lake, B. Divinskiy, G. Schmidt, S. O. Demokritov, and V. E. Demidov, "Interplay Between Nonlinear Spectral Shift and Nonlinear Damping of Spin Waves in Ultrathin Yttrium Iron Garnet Waveguides," Phys. Rev. Appl. **17**, 034010 (2022).

28. R. W. Boyd, *Nonlinear Optics. 4 Edition* (Academic Press, London, 2020).

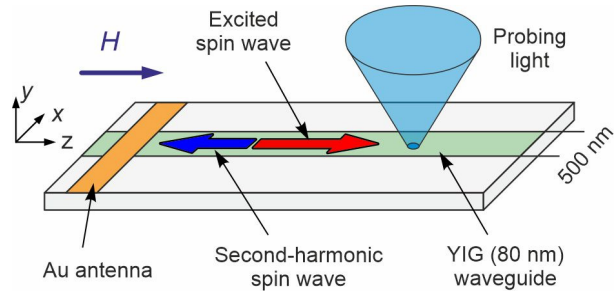

Fig. 1. Schematics of the experiment.

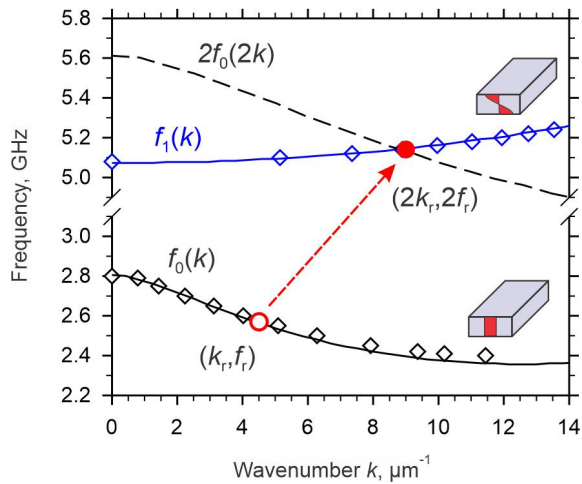

Fig. 2. Dispersion spectrum of spin waves in the studied waveguides. Diamonds show the experimental data. Curves labeled $f_0(k)$ and $f_1(k)$ show dispersion relations for the fundamental and the first-order thickness mode, respectively, obtained from the micromagnetic simulations. Insets schematically show the distributions of the dynamic magnetization for the two modes. Circles mark the spin-wave states, for which resonant conditions for the second harmonic generation are satisfied. The data are obtained at $H = 300$ Oe.

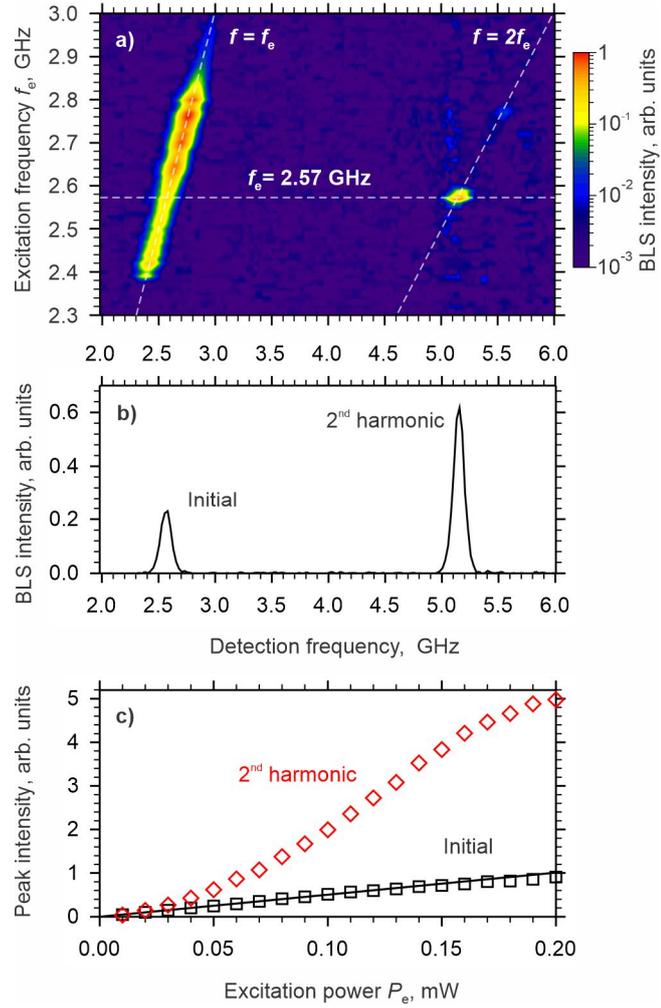

Fig. 3. (a) Color-coded map of the BLS intensity as a function of detection and excitation frequencies. Note the resonant excitation of the second harmonic close to $f_e$ = 2.57 GHz. (b) BLS spectrum recorded at the excitation frequency $f_e$ = 2.57 GHz (horizontal dashed line in (a)). The data are obtained at the power of the excitation signal $P_e$ = 0.05 mW. (c) Intensities of the second harmonic and the initial wave as a function of the excitation power. The data are obtained at the excitation frequency $f_e$ = 2.57 GHz. Line is the linear fit of the experimental data for the initial wave at $P_e$ < 0.15 mW. The data are obtained at $H$ = 500 Oe.

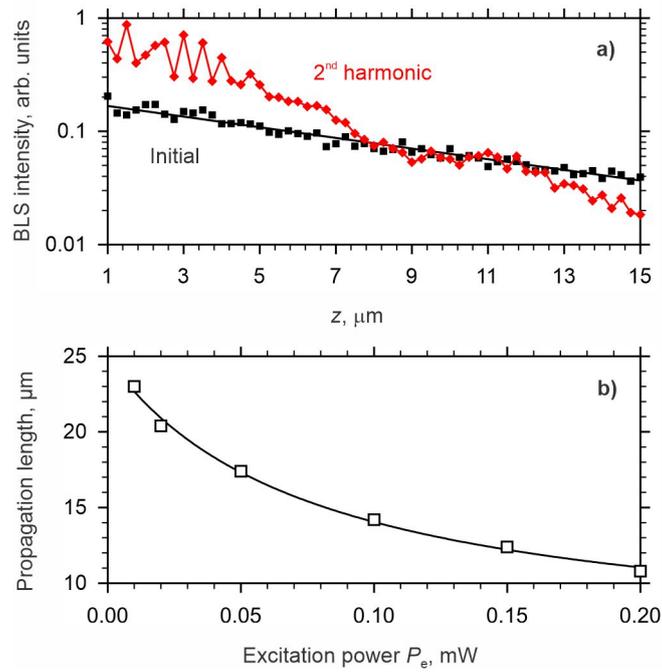

Fig. 4. (a) Spatial dependences of the intensities of the initial wave and the second harmonic. Line shows an exponential fit of the data for the initial wave (note the logarithmic scale of the vertical axis). (b) Power dependence of the propagation length of the initial wave. Symbols show the experimental data. Curve is a guide for the eye. The data are obtained at $H$ = 300 Oe.

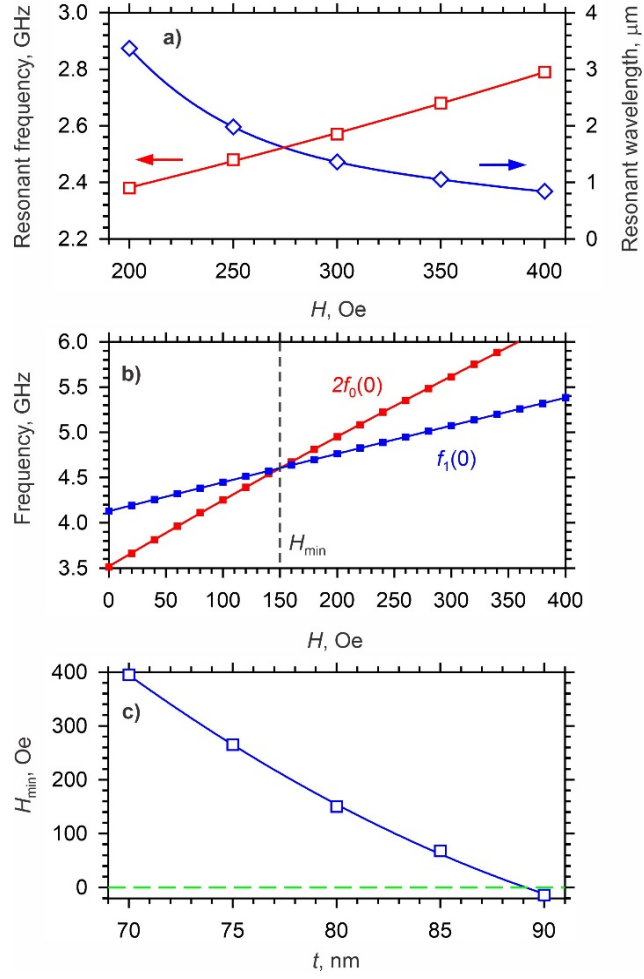

Fig. 5. (a) Dependences of the resonant frequency and the resonant wavelength on the static magnetic field. Symbols show the experimental data. Curves are guides for the eye. (b) Field dependences of the frequencies $2f_0(k=0)$ and $f_1(k=0)$ obtained from micromagnetic simulations. Vertical dashed line marks the minimum static magnetic field $H_{\min}$, at which the resonant conditions can be satisfied. (c) Dependence of the minimum field on the thickness of the waveguide obtained from micromagnetic simulations. Horizontal dashed line marks $H_{\min} = 0$. Curves in (b) and (c) are guides for the eye.